\documentclass[DIV=calc,onecolumn,preprintnumbers,amsmath,amssymb,superscriptaddress]{revtex4-2}

\usepackage{epsf}
\usepackage{graphicx}
\usepackage{sidecap}
\usepackage{amsmath}
\usepackage{float}
\usepackage[normalem]{ulem}
\usepackage{xcolor}

 \usepackage{stfloats}

\renewcommand{\thefigure}{{\textbf{\arabic{figure}}}}

\makeatletter
\def\maketitle{
\@author@finish
\title@column\titleblock@produce
\suppressfloats[t]}
\makeatother

\begin{document}

\title{Observation of Altermagnetic Spin-Splitting in an Intercalated Transition Metal Dichalcogenide}

\author{Milo Sprague}
\affiliation{Department of Physics, University of Central Florida, Orlando, Florida 32816, USA}
\author{Mazharul Islam Mondal}
\affiliation{Department of Physics, University of Central Florida, Orlando, Florida 32816, USA}
 \author{Anup Pradhan Sakhya}
 \affiliation{Department of Physics, University of Central Florida, Orlando, Florida 32816, USA}
 \affiliation{Research Institute for Synchrotron Radiation Science, Hiroshima University, Higashi-Hiroshima 739-0046, Japan}
 
\author{ Resham Babu Regmi}
\affiliation{Department of Physics and Astronomy, University of Notre Dame, Notre Dame, IN 46556, USA}
\affiliation{Stavropoulos Center for Complex Quantum Matter, University of Notre Dame, Notre Dame, IN 46556, USA}
\author{Surasree Sadhukhan}
\affiliation{Department of Physics and Astronomy, George Mason University, Fairfax, VA 22030, USA}
\affiliation{Quantum Science and Engineering Center, George Mason University, Fairfax, VA 22030, USA}
\author{Arun K. Kumay}
\affiliation{Department of Physics, University of Central Florida, Orlando, Florida 32816, USA}
\author{Himanshu Sheokand}
\affiliation{Department of Physics, University of Central Florida, Orlando, Florida 32816, USA}
\author{Igor I. Mazin}
\affiliation{Department of Physics and Astronomy, George Mason University, Fairfax, VA 22030, USA}
\affiliation{Quantum Science and Engineering Center, George Mason University, Fairfax, VA 22030, USA}
\author{Nirmal J. Ghimire}
\affiliation{Department of Physics and Astronomy, University of Notre Dame, Notre Dame, IN 46556, USA}
\affiliation{Stavropoulos Center for Complex Quantum Matter, University of Notre Dame, Notre Dame, IN 46556, USA}
\author{Madhab~Neupane}\email{e-mail: madhab.neupane@ucf.edu} 
 \affiliation{Department of Physics, University of Central Florida, Orlando, Florida 32816, USA}

\begin{abstract}

\centering\textbf{\large{Abstract}}

\justify Altermagnetism is a magnetic phase combining characteristics of both antiferromagnetism and ferromagnetic ordering. Despite growing theoretical interest in altermagnetic materials, reports of experimentally verified high-N\'eel temperature layered compounds are limited or remain to be firmly established. Here, we present an angle-resolved photoemission spectroscopy and density functional theory study of Co$_{1/4}$TaSe$_2$, a compound we identify as a layered altermagnetic material. Magnetic susceptibility measurements confirm type-A antiferromagnetic ordering with a Néel temperature of 178 K. Our spin-resolved and spin-integrated angle-resolved photoemission spectroscopy measurements reveal an electronic band structure in excellent agreement with density functional theory calculations, demonstrating clear signatures of altermagnetic spin splitting at the Fermi surface. Furthermore, temperature-dependent angle-resolved photoemission spectroscopy reveals a reconstructed valence band structure, with observable band shifts upon heating above the N\'eel temperature, consistent with the suppression of altermagnetic order. These findings establish Co$_{1/4}$TaSe$_2$ as a promising platform for exploring altermagnetic phenomena. 	\end{abstract}
\maketitle

\noindent\textbf{\large{Introduction}}\\

\noindent The interplay between magnetic ordering and electronic properties has been a prime topic of investigation within condensed matter physics since the field's inception. As a result of this work, magnetic materials have been implicated in emerging technologies, such as magnetic storage and spintronics \cite{Spintronics}. Long-ranged magnetic ordering has been traditionally classified into two main categories: ferro(ferri-)magnetic (FM) and antiferromagnetic (AFM). The distinguishing factor separating these phases is that, in the latter case, spin sublattices are related by a crystal symmetry that requires the net magnetization to be zero (and not compensated accidentally). An ideal FM possesses fully uncompensated magnetic moments, resulting in permanent net magnetization, which breaks time-reversal symmetry and lifts the spin-degeneracy of electronic states \cite{AHE_Nagosa_2010}.  Compensated AFM materials, on the other hand, feature vanishing net magnetic moments \cite{Neel_AFM}. The question of how compensated AFM ordering
 affects the electronic bands is highly contingent upon the crystallographic symmetries relating to magnetic sublattices. This distinction has led to the further classification of compensated collinear AFM materials under the term ``altermagnetism'', the altermagetic materials being such that (as opposed to compensated ferrimagnets) a crystal symmetry assures the compensation, but (as opposed to traditional antiferromagnets) this symmetry is neither translation nor inversion \cite{Punchline_Altermagnetism,Smejkal_Emerging_2022,SmejkalBeyond,Smejkal_AH_Antiferromagnets_2022,Magnetic_Group_Turek}. Altermagnets are distinguished from other compensated AFM materials by macroscopic electronic responses resembling ferromagnets, including spin-splitting of the electronic bands; however, just like in FMs, the exchange splitting is on the scale of the Hund's rule coupling (i.e., electronvolts), and, unlike FMs, it integrates to zero and shows a characteristic winding across momentum space \cite{Smejkal_Emerging_2022,SmejkalBeyond,FeSb2_Mazin, CrystalDesign_Wei,Chemical_Perspective_Fender}.

 The search for altermagnetic materials is driven by the need to overcome technological limitations inherent to both ferromagnets and antiferromagnets. In ferromagnetic spintronics, stray fields and spin torques introduce inefficiencies and crosstalk between devices, while antiferromagnets, despite their robustness against external perturbations, suffer from weak spin-readout signals due to the absence of a net magnetic moment \cite{Spintronics}. Altermagnets present a compelling alternative by combining the best aspects of both material classes: robust spin-polarized transport and the absence of stray fields. Early indications of altermagnetism came in compensated AFM materials showing the anomalous Hall effect \cite{Smejkal_AH_Antiferromagnets_2022}. Research into altermagnetic materials remains in its infancy, with limited experimental studies of altermagnetic band splitting being performed mostly on MnTe \cite{MnTe_Hariki_XMCD,MnTe_Kluczyk_AHE,MnTe_Lovesey_Templates,MnTe_Lee_ARPES,MnTe_Krempasky_ARPES,MnTe_Osumi_GiantBandSplitting,MnTe_Betancourt_AHE,MnTe_Mazin_Origin,MnTe_Regmi_Altermagnon} and CrSb \cite{CrSb_Reimers_ThinFilm,CrSb_Ding_gwave,CrSb_Zeng_ARPES}.

 Transition metal dichalcogenides (TMDs) have seen significant research interest due to their van der Waals layered structure and their exhibition of collective quantum phenomena, such as superconductivity, magnetism and charge density wave (CDW) formation \cite{TMDs_Wilson_1969,TMD_Electronic_Structure,TMD_Superconductivity_Theory_Neto_2001,NbSe2_CDW_FermiSurface_Rossnagel,NbSe2_CDW_FermiSurface_Straub,2H_TMDs_FermiSurface_Susceptibilities_Inosov, SmithARPESTaSe2_1985}. Due to the layered structure of these materials, intercalation of transition metal elements between the TMD layers has become a popular approach for tuning electronic and magnetic properties. The intercalated elements interact with the unique band structure of the TMD layers, introducing crystal symmetry modifications \cite{TMD_Intercalates_Marseglia_1983} and many-body instabilities, including emergent charge density wave ordering \cite{TMD_Intercalates_CDW_Baranov}, superconductivity \cite{TMD_Intercalates_Superconductivity_CuxTiSe2,TMD_Intercalates_Superconductivity_CuxTiSe2_DFT}, and itinerant magnetism \cite{TMD_Intercalates_Friend_1977,TMD_Intercalates_Magnetic_1980,TMD_Intercalates_Marseglia_1983}. While the choice of intercalant is a key design factor, the intercalated sublattice structure also provides significant control. Recently, the intercalated TMD compound, Co$_{1/4}$NbSe$_2$, has been studied, which contains cobalt atoms forming a hexagonal sublattice between van der Waals layers of 2H-NbSe$_2$, preserving the space group of the parent compound, including inversion symmetry \cite{CoNbSe_Regmi_Intro,CoNbSe_Sakhya, CoNbSe_muSR, CoNbSe_gwaveARPES}. Neutron diffraction, magnetic susceptibility \cite{CoNbSe_Regmi_Intro}, and $\mu$SR measurements \cite{CoNbSe_muSR} were used to characterize the magnetic behavior of Co$_{1/4}$NbSe$_2$, which revealed a type-A AFM ordering with moments oriented along the $c$-axis.  Density functional theory (DFT) calculations and angle-resolved photoemission spectroscopy (ARPES) measurements \cite{CoNbSe_Sakhya,CoNbSe_gwaveARPES} indeed revealed the presence of altermagnetic splitting along the $\Gamma' - \text{M}'$ direction within the $k_z$ = $\pi$/2$c$ plane (See Fig. 1c).

We synthesized Co$_{1/4}$TaSe$_2$ by intercalating cobalt, which uniformly doubles the in-plane 2H-TaSe$_2$ unit cell (Fig. 1a,b). Our magnetic characterization suggests that the Co atoms order antiferromagnetically below 178 K with an easy-axis anisotropy, consistent with previous reports of a type-A AFM via neutron scattering \cite{MandujanoCoTaSeMagnetic}. 

 The crystal and magnetic structure shows that the AFM symmetry contains a mirror plane, while the opposite spin-sites are not related by a simple inversion or translation, thus satisfying the condition of altermagnetism. Compared to Co$_{1/4}$NbSe$_2$, Co$_{1/4}$TaSe$_2$ brings more extended 5$d$ orbitals, leading to higher Fermi velocities, modified $k_z$ dispersion, and smaller net spin polarization of the Fermi surface for comparable energetic splitting. These differences place Co$_{1/4}$TaSe$_2$ in a distinct physical regime, allowing direct tests of how altermagnetism scales with bandwidth and dimensionality.
  Recent neutron diffraction studies reported a local Co moment of 1.35 $\mu_{\text{B}}$, lower than the expected 3 $\mu_{\text{B}}$ in the high-spin $d^7$ configuration, suggesting involvement of Co bands in the valence band structure \cite{MandujanoCoTaSeMagnetic}. Combined with reduced magnetic entropy recovery, this suggests partial delocalization of the magnetic moments, likely involving hybridization with Ta 5$d$ orbitals. The partial delocalization of Co moments, typical of 3$d$ transition metals, combined with local crystal environments favoring altermagnetism, strongly motivates an electronic structure study of Co$_{1/4}$TaSe$_2$.

In this work, we combine spin-resolved and spin-integrated ARPES measurements with DFT calculations to investigate the valence band structure of Co$_{1/4}$TaSe$_2$. Our results indicate signatures of altermagnetic band splitting in this material.\\

\noindent\textbf{\large{Results}}\\

\noindent\textbf{C\lowercase{o}$_{1/4}$T\lowercase{a}S\lowercase{e}$_2$ Crystal Structure and Altermagnetic Characteristics}

\begin{figure*}[!htbp]
	\centering
	\includegraphics[width=1\linewidth]{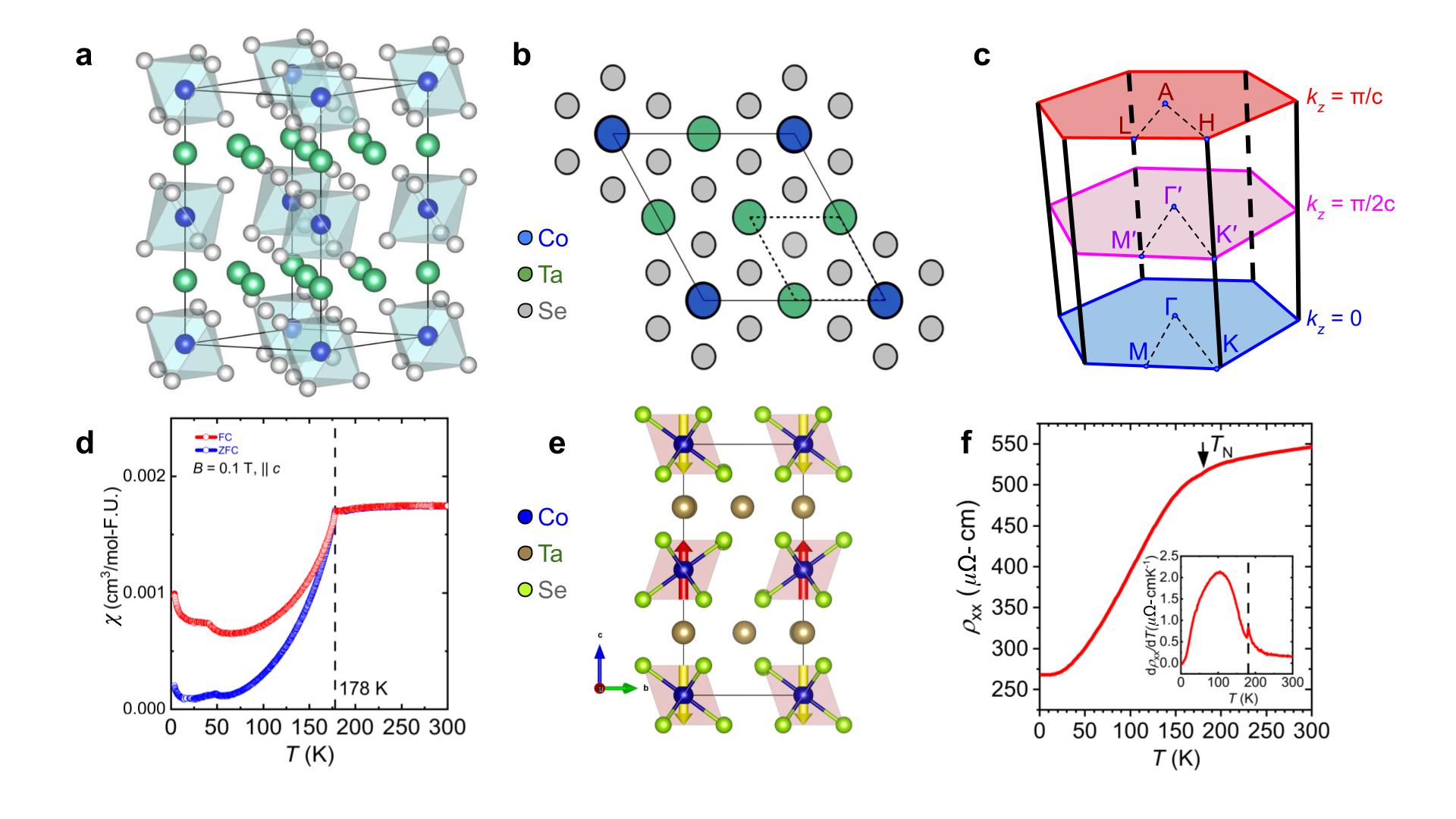}
	\caption{\textbf{Establishment of Altermagnetic Crystal and Magnetic Structure in Co$_{1/4}$TaSe$_2$.} \textbf{a} 3D visualization of the Co$_{1/4}$TaSe$_2$ unit cell, with Co, Ta, and Se atoms being represented by blue, green, and gray spheres, respectively. \textbf{b} Viewing of the Co$_{1/4}$TaSe$_2$ crystal structure along the $c$-axis. The smaller dashed parallelogram represents the 1H-TaSe$_2$ unit cell, and the larger solid one represents the Co$_{1/4}$TaSe$_2$ unit cell. \textbf{c} Bulk hexagonal Brillouin zone with high-symmetry points labeled. \textbf{d} Magnetic susceptibility of Co$_{1/4}$TaSe$_2$ featuring a clear magnetic transition at $T_\text{N}$=178 K, as indicated by the dashed vertical line. The field-cooled (FC) and zero field-cooled (ZFC) susceptibilities were measured under a 0.1 T field applied parallel to the $c$-axis and are presented as red and blue points, respectively. \textbf{e} Magnetic configuration with spin up (red) and spin down (yellow) pointing along the $c$-axis. Moments are ordered ferromagnetically in the plane and antiferromagnetically between $c$-axis layers. \textbf{f} Longitudinal resistivity of Co$_{1/4}$TaSe$_2$ single crystals as a function of temperature. The N\'eel temperature at $T_{\text{N}}$ = 178 K is indicated by the downward-pointing arrow at the top of the panel. The bottom-right inset presents the derivative of the resistivity with temperature with $T_{\text{N}}$ indicated by the vertical dashed line.}
	\label{fig:cotasefigure2}
\end{figure*}

\noindent Co$_{1/4}$TaSe$_2$ crystallizes in the hexagonal P6$_3$/mmc space group (No. 194), matching that of the 2H-TaSe$_2$ polytype. The experimental lattice constants are a = 6.8828(1) $\text{\AA}$ and c = 12.4535(3) $\text{\AA}$ \cite{MandujanoCoTaSeMagnetic}. The structure within the unit cell consists of 1H-TaSe$_2$ layers separated by intercalated Co atoms, as shown in Fig. 1a. The green, grey, and blue spheres represent Ta, Se, and Co atoms, respectively. These cobalt atoms occupy the 2a Wyckoff site, which preserves the space group and introduces a 2$\times$2 doubling of the 2H-TaSe$_2$ unit cell within the $ab$-plane, which can be seen from the $c$-axis view, as depicted in Fig. 1b. 

The 2H-TaSe$_2$ unit cell is shown as the smaller dashed line while the Co$_{1/4}$TaSe$_2$ unit cell is outlined by the solid line. Fig. 1c shows the hexagonal bulk BZ, with the $k_z$ = 0, $\pi/2c$, and $\pi/c$ planes marked in blue, pink, and red, respectively. High-symmetry points within these planes are labeled as $\Gamma$, M, K ($k_z$ = 0); $\Gamma'$, M$'$, K$'$ ($\pi/2c$); and A, L, H ($\pi/c$).

The magnetic character of Co$_{1/4}$TaSe$_2$ sample was established using magnetic susceptibility measurements, shown in Fig. 1d. We find a sharp suppression of the magnetic susceptibility below 178 K, indicating antiferromagnetic ordering below this temperature \cite{MandujanoCoTaSeMagnetic}. This transition temperature is further corroborated by temperature-dependent heat capacity measurements, which show a clear lambda-like anomaly at $T_{\text{N}}$ = 178 K (See Supplementary Fig. 1). Figure 1(e) presents a diagram that depicts the magnetic configuration within the unit cell. The moments are centered on the Co 3$d$ orbitals and show type-A behavior, with ferromagnetic in-plane exchange interactions and AFM out-of-plane coupling. 

Our DFT calculations show that magnetic moments are well localized on Co, with only about 10\% of the opposite-sign magnetization generated in the TaSe$_2$ layers, indicating that the reduction of the moment compared to ideal Co$^{2+}$ high-spin state is due to Co forming an itinerant band, corresponding to the non-integer valence state. Indeed, recent neutron scattering studies on Co$_{1/4}$TaSe$_2$ report a reduced magnetic moment on the Co sites, but no magnetization on other atoms \cite{MandujanoCoTaSeMagnetic}.

Fig. 1f presents the in-plane longitudinal resistivity ($\rho_{xx}$) as a function of temperature and the temperature-derivative of the resistivity ($d\rho_{xx}/dT$) within the Fig. 1f inset. The resistivity increases monotonically with temperature, which suggests metallic behavior. The temperature variation of the resistivity is much larger below the N\'eel transition, which is common in materials with strong scattering of carriers off spin fluctuations. This influence of the magnetic ordering on the resistivity suggests that Co electrons are itinerant and contribute substantially to overall conductivity. Indeed, our DFT calculations suggest that about a quarter of the total density of states at the Fermi level comes from Co. The magnetic and transport properties of our samples are consistent with the previous report \cite{MandujanoCoTaSeMagnetic}.\\

\noindent\textbf{Observation of Altermagnetic Spin-Splitting}\\

\begin{figure*}[!h]
	\centering
	\includegraphics[width=1\linewidth]{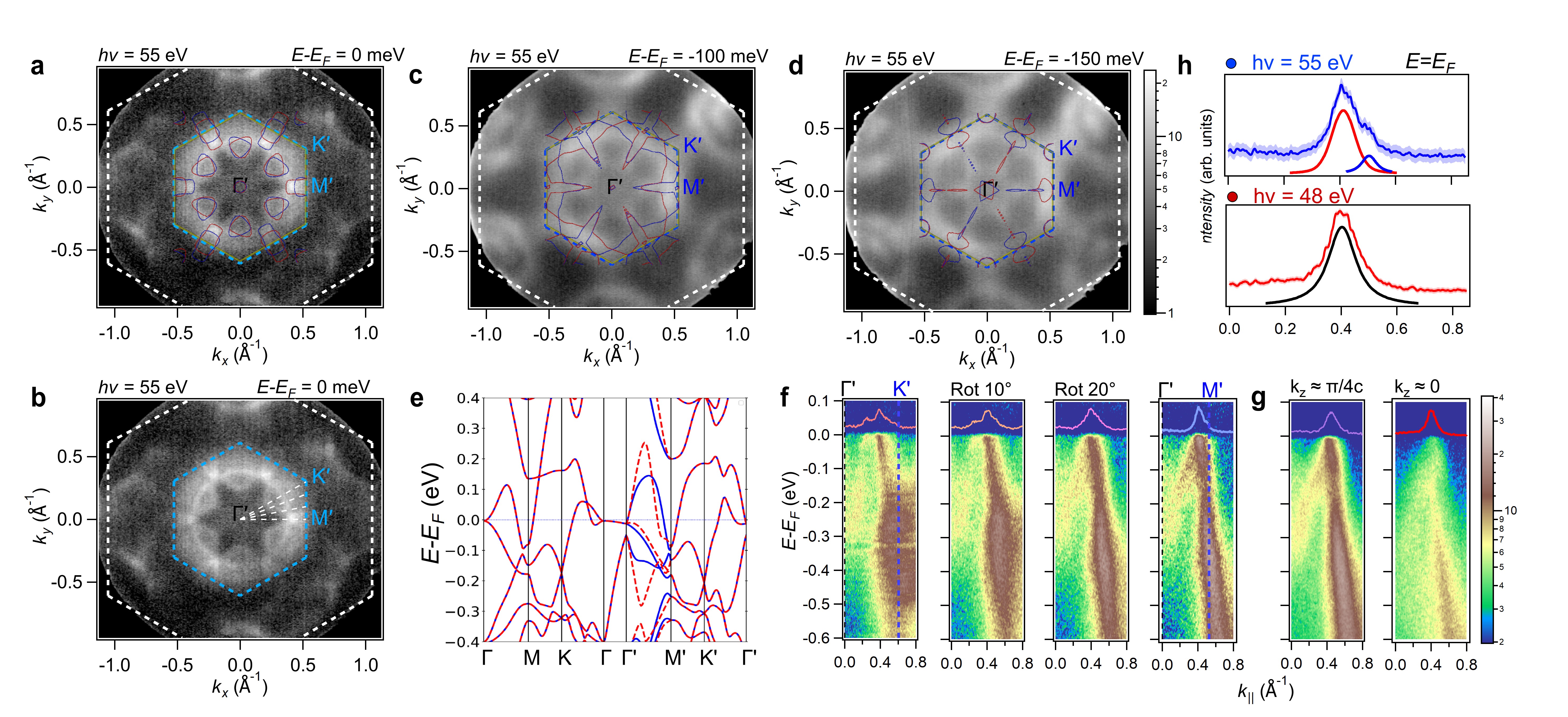}
	\caption{\textbf{Observation of altermagnetic spin-splitting in Co$_{1/4}$TaSe$_2$}. \textbf{a,b} ARPES-measured Fermi surface with and without the DFT-calculated Fermi surface overlaid. \textbf{c,d} Further CECs taken at $E-E_F$ = -100 meV and -150 meV, respectively. \textbf{e} Calculated electronic band dispersion taken along several high-symmetry directions. Unprimed high-symmetry labels denote points residing within the $k_z$ = 0 plane, while primed labels correspond to points lying in the $k_z$ = $\pi/2c$ plane. The red dashed and solid blue lines denote opposite-spin bands, which show spin-degeneracy along every presented direction except $\Gamma'$-$\text{M}'$. \textbf{f} ARPES-measured band dispersion along $\Gamma'-\text{K}'$ (left-most panel), then rotated by 10$^{\circ}$ increments toward the $\Gamma'-\text{M}'$ direction. MDCs taken along each of the cuts at the Fermi energy are shown on the top of each panel, with colors ranging from pink ($\Gamma'-\text{K}'$) to blue ($\Gamma'-\text{M}'$) corresponding to the rotation angle. \textbf{g} Decreasing $k_z$ from $\pi$/2$c$ toward $k_z\approx$ 0 ($\Gamma-\text{M}$). The purple and red MDCs are taken at the Fermi energy and correspond to $k_z\approx\pi$/2$c$ and 0, respectively. \textbf{h} The MDC taken at the Fermi energy along the $\Gamma'-\text{M}'$ direction measured using 55 eV photons (top panel, in blue) and 48 eV (bottom panel, in red). The shaded region surrounding each MDC indicates the standard deviation of photoemission counts measured for MDCs above the Fermi energy. Curve fitting using Voigt functions was performed to confirm the size of the altermagnetic spin-splitting and estimate the lineshape of the individual bands. The sample temperature was maintained at 7 K, well into the altermagnetic phase. ARPES measurements were performed at SSRL Beamline 5-2.}
	\label{fig:cotasefigure2}
\end{figure*}

\noindent Given the support of altermagnetism in
the crystal and magnetic structure, we search for evidence of spin-splitting in the electronic bands. ARPES is widely considered to be the most direct experimental probe of the electronic band structure, which motivates our use of it alongside DFT calculations. The sample temperature was maintained at T = 7 K, well into the AFM phase. Core-level photoemission reveals distinct Ta 4$f$, Ta 5$p$, and Se 3$d$ peaks. Notably, the Se peaks are accompanied by two shallower features, which can be attributed to shifted orbital energies near the surface, consistent with a Se cleavage plane \cite{Vita_Core-Level}(see Supplementary Fig. 2). Figure 2a reveals the Fermi surface measured using a photon energy of 55 eV and calculated within the $k_z$=$\pi/2c$ plane (where theoretically the maximal splitting is expected). Calculations are overlaid upon the experimental Fermi surface, with red/blue lines indicating the spin-up/down Fermi surfaces. For comparison, the experimental ARPES measurement without the calculations overlaid is shown in Fig. 2b. The Fermi surface is described by a small electron-like pocket surrounding the $\Gamma'$ point, petals along the $\Gamma'-\text{K}'$ direction, and the dog-bone pockets surrounding the $\text{M}'$ point.

The spin-splitting of the Fermi surface shows clear $g$-wave symmetry with a six-fold alternation of the opposite-spin Fermi surfaces. The Fermi surfaces show spin-degeneracies along the $\Gamma'-\text{K}'$ directions. Moving away from the $\Gamma'-\text{K}'$ direction, the magnitude of the splitting increases when rotating toward the $\Gamma'-\text{M}'$ direction. The altermagnetic spin-splitting is most prominent on the dog-bone pocket, showing clear separation of the opposite-spin Fermi surfaces along $\Gamma'-\text{M}'$. The evolution of the Fermi pockets with increasing binding energy is shown in Figs. 2c,d for $E-E_{F} = -100$ meV and $-150$ meV, respectively. In Fig. 2c, the closed petal pockets along the $\Gamma'-\text{K}'$ direction transform into open spin-split pockets crossing the $\text{K}'$ point. Continuing down to Fig. 2d, the Fermi surface shows a highly spin-polarized constant energy contour, with both the narrow elliptical pockets and the fan-shaped pockets along the $\Gamma'-\text{M}'$ direction showing complete spin-polarization.

 Next, we turn our attention to the electronic band dispersion. The electronic bands are calculated between the three-dimensional high-symmetry directions within the $k_z = 0$ and $\pi/2c$ planes, with high-symmetry points labeled according to their $k_z$ plane with unprimed and primed labels, respectively, as shown in Fig. 2e. Complete spin-degeneracy of the electronic bands are observed except along the $\Gamma'-\text{M}'$ direction, where the separation between the solid blue line and dashed red line represents the splitting between opposite-spin electrons. The dog-bone pocket can be identified as the two steeply dispersing bands crossing the Fermi energy close to the M$'$ point. The momentum separation between these Fermi crossings is about 0.01 $\text{\AA}^{-1}$. We compare our DFT calculation with the ARPES-measured dispersion along the $\Gamma'-\text{K}'$ and for angles rotated by 10$^{\circ}$, 20$^{\circ}$, and 30$^{\circ}$ toward $\Gamma'-\text{M}'$ in the four panels of Fig. 2f. The directions of these cuts are also shown in Fig. 2b for clarity. The two right panels in Fig. 2g demonstrate the dispersion decreasing the photon energy from $h\nu$ = 55 eV ($k_z\approx\pi/2c$) to 51 eV ($k_z\approx\pi/4c$) and 48 eV ($k_z \approx 0$), respectively. The three-dimensional dispersion of the low-energy bands is more clearly presented in Supplementary Fig. 5. The $\Gamma'-\text{K}'$ dispersion shows two spectral features crossing the Fermi energy, which forms the front (left) and back (right) of the petal Fermi pockets. Rotating toward the $\Gamma'-\text{M}'$ direction shows a gradual broadening of the spectral intensity corresponding to the gradual separation of the spin-split bands. Similarly, moving away from the $k_z = \pi/2c$ plane reduces the broadness of the spectral feature following the reduction of the spin-splitting. The momentum distribution curves (MDCs) integrated over a range of $\pm13$ meV were taken at the Fermi energy and are indicated in the right-most panels of Figs. 2f,g. 
 
 A zoomed-in view of these MDCs are shown in Fig. 2h, where the solid line represents the raw photoemission intensity and the shaded area surrounding the solid line represents the standard deviation of the noise taken above the Fermi energy. The MDC taken at $h\nu$ = 55 eV shows a clear hump on the right side, indicating the presence of a second feature with a separation close to the momentum resolution limit. We perform a fit of the MDC using Voigt functions, where we find two peaks of similar width separated by $\sim 0.09 ~\text{\AA}^{-1}$, close to the instrumental (Gaussian) peak width of $\sim 0.08 ~\text{\AA}^{-1}$. This is to be contrasted with the MDC obtained along the $\Gamma-\text{M}$ using $h\nu$ = 48 eV, which shows a single-peak lineshape.\\

\noindent\textbf{Electronic Band Structure Across the Altermagnetic Transition}\\

\begin{figure*}[h!]
	\centering
	\includegraphics[width=1\linewidth]{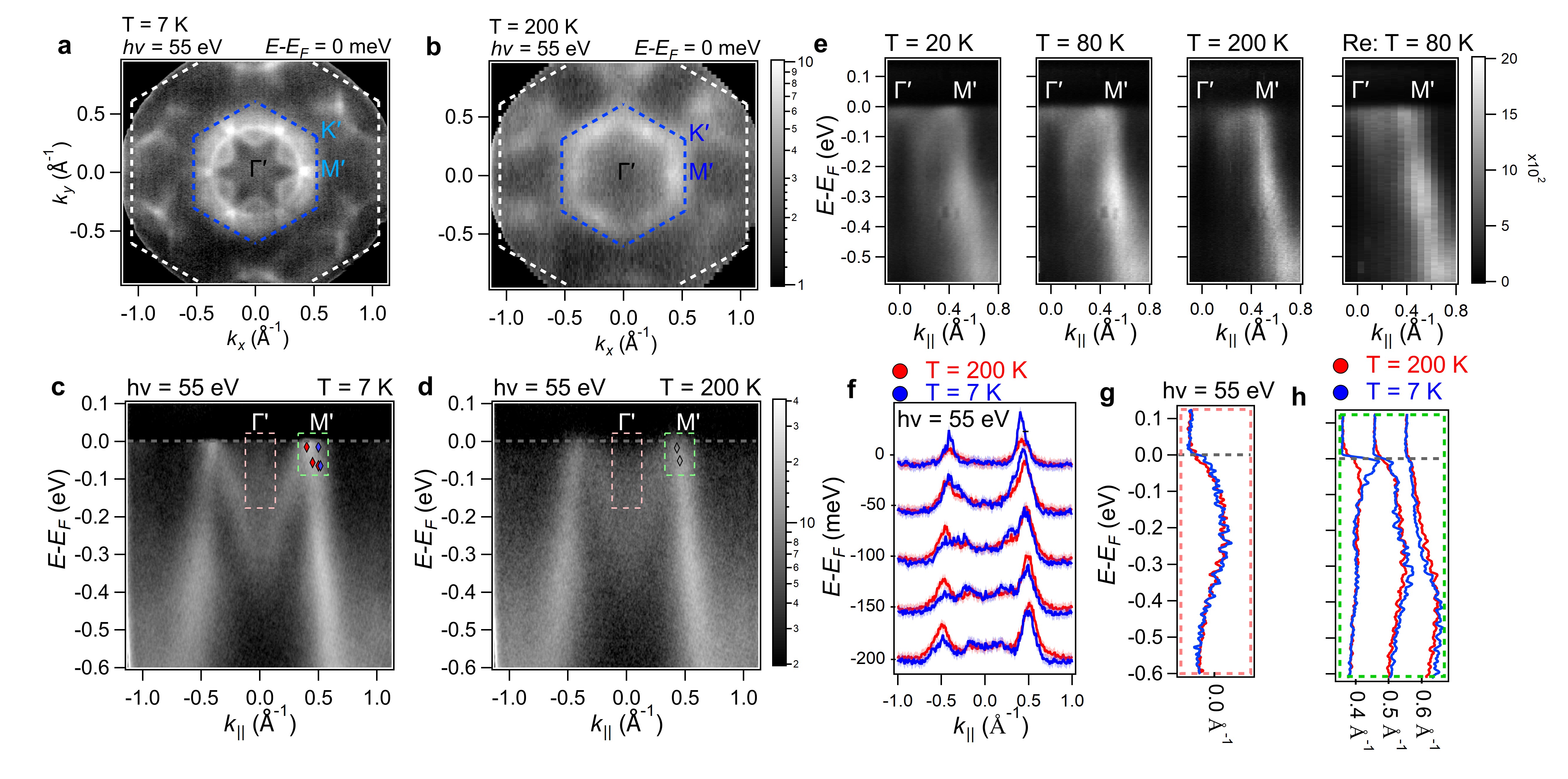}
	\caption{\textbf{Temperature evolution of the Electronic Structure across $T_{\text{N}}$.} \textbf{a,b} ARPES-measured Fermi surface measured using 55 eV photons at a sample temperature of 7 K and 200 K, respectively. \textbf{c,d} Corresponding band dispersion along the $\Gamma'-\text{M}'$ for low and high temperatures. The red and green boxes highlight parts of the dispersion spectrum showing clear modifications with temperature. \textbf{e} The temperature variation of the band dispersion along the $\Gamma'-\text{M}'$ direction at T = 20 K, 80 K, 200 K, and cycled back to 80 K. \textbf{f} Stacked MDCs along the $\Gamma'-\text{M}'$ direction integrated over 15 meV and offset by 50 meV. The shaded region surrounding each MDC indicates the standard deviation of photoemission counts measured for MDCs above the Fermi energy. \textbf{g,h} Experimental EDCs stacked along the $\Gamma'-\text{M}'$ direction taken from within the red and green boxes.}
	\label{fig:layout2}
\end{figure*}

\noindent The influence of magnetic ordering on the electronic band structure is further investigated by varying the temperature across $T_{\text{N}}$ = 178 K. Figures 3a,b present the ARPES-obtained Fermi surfaces, measured using 55 eV photons, at a temperature of 7 K and 200 K, respectively. The Fermi surface shows a significant broadening of its features, which produces an enlargement of the Fermi pockets along the $\Gamma'-\text{K}'$ direction. Similarly, the intensity coming from the petal-shaped pockets along $\Gamma'-\text{K}'$ is significantly reduced above $T_{\text{N}}$, however, the band forming this feature doesn't disappear, as it can still be resolved in the dispersion cuts across both temperatures (See Supplementary Fig. 3). Figs. 3c,d focus on the dispersion cuts along the $\Gamma'-\text{M}'$ direction. Attention is drawn to two regions shown within the red and green boxes, which contain the most apparent modifications to the spectra. The red box in Fig. 3c shows the band forming the small hole-like pocket around the $\Gamma'$ point. Compared with the high-temperature result within the same region in Fig. 3d shows the suppression of the spectral intensity at the Fermi energy, potentially indicating a downward shift of the pocket away from $E_F$. The region within the green box highlights the spin-split bands producing the dog-bone pocket discussed in Fig. 2. The low-temperature cut shows a two-peaked spectrum corresponding to the altermagnetic splitting (red and blue diamonds in Fig. 2c), while the high-temperature spectrum shows a narrower band crossing the Fermi energy (gray diamonds in Fig. 2d). Measurements at intermediate temperatures between 20 K and 80 K, presented in Fig. 3e, establish that both the hole-like band around the $\Gamma'$ point and the widened dog-bone feature persist when heating the sample to 20 K and 80 K, but disappears for T = 200 K, above $T_\text{N}$. Cycling the temperature back down below $T_\text{N}$ reestablishes the low-temperature features. The dispersion cuts provide an overview of the temperature-dependent modifications, though consideration of the energy distribution curves (EDCs) and momentum distribution curves (MDCs) provides a more direct comparison of the photoemission intensities. Fig. 3f presents stacked MDCs integrated over a 15 meV window and taken 50 meV apart. The MDC peak positions exhibit minimal variation between the T = 7 K and T = 200 K profiles, suggesting that temperature-dependent modifications to the band structure are comparable to or smaller than the experimental resolution. Differences in the MDC peak intensities provide clues as to how these details of the band structure are modified across the magnetic transition. The innermost peaks display reduced intensity at T = 200 K across all binding energies considered, consistent with thermal broadening of these features. The outer peaks, however, retain similar intensities for -100 meV $<E-E_F<$ 0 meV, but become suppressed in the range -200 meV $<E-E_F<$ -100 meV. This behavior may reflect the closing of small altermagnetism-induced band gaps at elevated temperatures. Stacked EDCs taken within the corresponding red (green) boxes in Figs. 3c,d are shown in Fig. 3g (Fig. 3h). The blue (red) EDC lines provide the 7 K (200 K) sample temperature results. 

The altermagnetic spin-split bands discussed in Fig. 2 show a reduced spectral intensity in the PM phase, as is seen in the left-most EDC in Fig. 2f. This effect may arise due to changing photoemission matrix element effects with modified orbital composition across the magnetic transition. In addition to this reduction near the Fermi energy, further redistribution of the spectral intensity is observed within the remaining EDCs of Fig. 2f, which are likely due to a combination of thermal broadening, suppression of altermagnetic splitting, and magnetic gaps closing near the $\text{M}'$ point.\\

\noindent\textbf{Spin Texture of the Altermagnetic Band Structure}\\

\begin{figure*}
    \centering
    \includegraphics[width=1\linewidth]{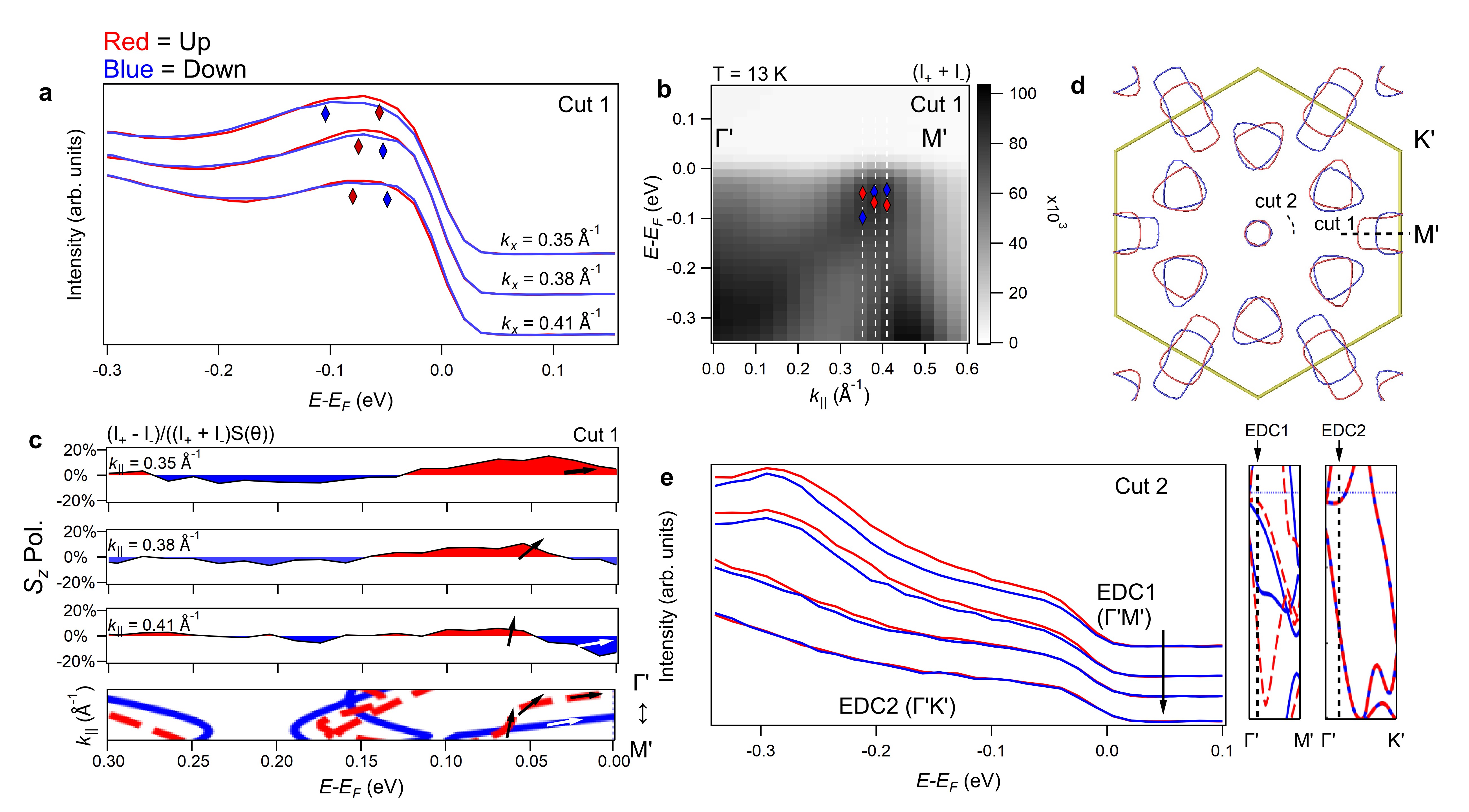}
    \caption{\textbf{Spin-polarization of electronic bands via Spin-ARPES.} \textbf{a} Spin-resolved EDCs corresponding to the momenta indicated in (b). The relative energy position of the spin-up and spin-down peaks shift continuously across the dog-bone pocket, demonstrating a reversal of the spin polarization and a finite energy splitting between the two spin channels. \textbf{b} Band dispersion from the total photoemission intensity ($I_++I_-$) along the $\Gamma'-\text{M}'$ direction across the dog-bone pocket. The three vertical dashed lines indicate the momenta at which spin-resolved EDCs are extracted ($k_{||} = 0.41,\ 0.38,\ \text{and}\ 0.35\ \text{\AA}^{-1}$). Colored markers denote the approximate peak positions of the spin-up (red) and spin-down (blue) components, illustrating the momentum-dependent reversal of the dominant spin character across the band. \textbf{c} The out-of-plane spin polarization ($\langle S_z\rangle$) for the EDCs presented in \textbf{a} using the measured asymmetry divided by the Sherman function. The polarization exhibits a clear sign reversal across the dog-bone band, varying from approximately -13$\%$ to +13$\%$. The bottom panel depicts the calculated electronic band dispersions over that same interval. Arrows serve as a guide to the eye for corresponding the calculated dispersions with the observed evolution of the spin polarization. \textbf{d} DFT calculated Fermi surface with the relevant cuts overlaid. \textbf{e} The spin-resolved EDCs taken along a curved path between a point on the $\Gamma'-\text{M}'$ line (EDC1) and the $\Gamma'-\text{K}'$ direction (EDC2). As the EDCs are rotated, the spin polarization continuously decreases until it reaches the nodal plane along the $\Gamma'-\text{K}'$ direction. The calculated band dispersions along the $\Gamma'-\text{M}'$ and $\Gamma'-\text{K}'$ directions are reproduced in the right two panels to illustrate the end points of the path. Vertical dashed lines indicate the momenta locations of EDC1 and EDC2. \textbf{(a-c)} were taken along cut 1, along the $\Gamma'-\text{M}'$ direction, while panel \textbf{e} was taken along the curved path, cut 2.}
    \label{fig:placeholder}
\end{figure*}
\noindent To assess the origin of the observed band splitting in the ARPES spectrum, we performed spin-resolved ARPES using a Scienta DA30 hemispherical analyzer equipped with a 3D Spin VLEED detector, focusing our attention on the out-of-plane $S_z$ spin component (see Fig. 1e). We consider the EDCs across the dog-bone pocket along the $\Gamma' - \text{M}'$ direction in Figures 4a,b. Fig. 4a presents the spin-resolved EDCs obtained during spin-resolved ARPES measurements taken across the dog-bone pocket along the $\Gamma'-\text{M}'$ direction, taken at $k_{||} =$ 0.35 $\text{\AA}^{-1}$, 0.38 $\text{\AA}^{-1}$, and 0.41 $\text{\AA}^{-1}$. For ease of interpretation, Fig. 4b presents the spin-integrated ARPES intensity with approximated spin-up (down) EDC peak positions indicated by red (blue) diamonds, capturing the independent dispersion of these two spin channels. Across these selected momenta, the spin-resolved EDCs reveal a clear evolution of the spectral weight associated with opposite spin channels. At $k_{||}=0.41~\text{\AA}^{-1}$, the spin-down component exhibits a peak shifted closer to the Fermi energy relative to the spin-up component, indicating an energy splitting between the two spin levels. Moving toward $k_{||}=0.38~\text{\AA}^{-1}$, the spectral features corresponding to the two spin channels become more closely aligned in energy. On the opposite side, at $k_{||}=0.35~\text{\AA}^{-1}$, the spin-down peak shifts downward toward higher binding energy and the spin-up component moves closer to the Fermi energy. This evolution demonstrates a flipping of the dominant spin character across the dog-bone pocket, in agreement with the spin texture predicted by our DFT calculations.\\
\indent The corresponding spin polarization extracted from the EDCs is shown in Fig. 4c. The polarization is obtained from the asymmetry between the spin-resolved intensities, normalized by the Sherman function of the detector. Across the dog-bone pocket, the low-energy spin polarization exhibits a clear sign change, evolving from positive to negative as momentum is swept across this spectral feature. This sign reversal is consistent with the momentum-dependent spin-splitting described above. To further scrutinize the g-wave character of the altermagnetic splitting, we present the EDCs taken along the curved path, connecting a point along the $\Gamma'-\text{M}'$ direction, where pronounced altermagnetic splitting is expected, with a point along the $\Gamma'-\text{K}'$ direction, where spin-degeneracy is expected. For clarity, we have included a diagram of the cut directions in Fig. 4d, where cut 1 is along the $\Gamma'-\text{M}'$ line and cut 2 is the curved trajectory. Indeed, the EDCs taken along cut 2 are presented in Fig. 4e, showing a decreasing spin polarization when moving from the $\Gamma'-\text{M}'$ line to the $\Gamma'-\text{K}'$ line. Taken together, our spin-resolved ARPES measurements establish that the electronic states near the Fermi level not only show clear spin-polarization, but also carry a well-defined and alternating $S_z$ polarization, consistent with the predicted altermagnetic ground state.\\

\indent It is important to note that the two cobalt sites within the unit cell are related by mirror symmetry and a general $\mathit{k}$ does not retain Kramers degeneracy. However, the hexagonal crystal symmetry forces nodal (spin-degenerate) planes within the $\Gamma$-K-H-A plane, along with the $k_z=0$ and $k_z=\pi/c$ planes (see Fig. 2).
Therefore, the electronic structure obtained in DFT calculations and confirmed via ARPES measurements, as presented in Fig. 2, exhibits clear \textit{g}-wave splitting. Previous reports on MnTe and CrSb similarly revealed $g$-wave altermagnetic splitting \cite{MnTe_Hariki_XMCD,MnTe_Kluczyk_AHE,MnTe_Lovesey_Templates,MnTe_Lee_ARPES,MnTe_Krempasky_ARPES,MnTe_Osumi_GiantBandSplitting,MnTe_Betancourt_AHE,MnTe_Mazin_Origin,MnTe_Regmi_Altermagnon,CrSb_Reimers_ThinFilm,CrSb_Ding_gwave,CrSb_Zeng_ARPES}. Co$_{1/4}$TaSe$_2$ is unique among the discussed altermagnetic materials due to its layered van-der Waals structure, opening a pathway for potential implementation in heterostructures and interfaces involving topological materials and superconductors \cite{Applications_Andreev_Reflection}. Recent studies have extended this concept to engineered systems, including epitaxial thin films \cite{Santhosh_CrSb_Film}, few-layer altermagnetic bilayers exhibiting all-electrical layer-spin control \cite{Peng_Bilayer}, and heterostructures enabling electrical manipulation of spin transport \cite{Fu_Heterostructures}. Within this emerging context, an altermagnetic intercalated transition metal dichalcogenide such as Co$_{1/4}$TaSe$_2$ provides an intrinsically layered, van der Waals platform that bridges bulk and interface-based altermagnetism, positioning it as a promising candidate for future device-oriented and heterostructure studies. Furthermore, there is currently significant research interest in synthesizing Moir\'e lattices in transition-metal dichalcogenides.

Taken together, these finding support a dual character of magnetism in these Co-intercalated 2H-TMD systems. The Co moments exhibit partial localization, which allows for RKKY-mediated coupling to the itinerant electrons and results in band hybridization features appearing below $T_{\text{N}}$. Simultaneously, the itinerant component is responsible for momentum-selective gap openings and redistributions of spectral weight at the Fermi level, consistent with the direct participation of the Fermi surface in the magnetic ordering. This intermediate picture, which bridges the physics of localized and itinerant limits, aligns with broader observations in other 3$d$ transition metal systems where incomplete moment recovery and momentum-dependent reconstructions coexist. This interplay leaves a clear imprint on the low-energy electronic structure. The absence of strong precursor fluctuations above $T_{\text{N}}$ suggests that this interplay becomes operative only in the magnetic ordered state, pointing to a cooperative mechanism in which static order locks in the reconstruction of the electronic bands.

We have revealed a characterization of the magnetic and electronic structure of Co$_{1/4}$TaSe$_2$. This is an intercalated TMD compound exhibiting a hexagonal symmetry of magnetization in the momentum space, i.e., a g-wave altermagnet. Low-temperature spin-resolved and spin-integrated APRES is used in conjunction with first-principles DFT calculations to study the altermagnetic electronic structure, where a clear g-wave spin-splitting of the Fermi surface and valence band structure is observed in agreement with the calculated spin-resolved band structure. This spin splitting is present when the electronic structure is measured using 55 eV photons, corresponding to the $k_z\approx \pi/2c$ plane, whereas measuring the same band using 48 eV photons ($k_z\approx 0$) shows a vanishing of the splitting. The ARPES-measured band structures are compared below (T = 7 K) and above (T = 200 K) the N\'eel temperature ($T_{\text{N}}$ = 178 K). In addition to the vanishing of the altermagnetic spin-splitting for the high-temperature, several more dramatic reconstructions of the band structure are observed, including the closing of the small $\Gamma'$ pocket and the reduction in the spectral intensity at the Fermi surface. This work, in conjunction with recent $\mu$SR and neutron scattering studies of the magnetic structure, suggests an intermediate localized-itinerant behavior of the moments, based on observed modifications of the electronic structure both near and away from the Fermi energy. Thus, Co$_{1/4}$TaSe$_2$ is a promising platform for the manifestation of altermagnetism as an additional factor in understanding the complex interplay of magnetism and the electronic band structure of a material.\\

\noindent\large{\textbf{Methods}}\\

\noindent\normalsize\textbf{Crystal growth and characterization}\\ 

\noindent Co$_{1/4}$TaSe$_2$ single crystals were synthesized via chemical vapor transport using iodine as the transport medium. Initially, a polycrystalline precursor was prepared by sealing stoichiometric amounts of cobalt powder (Alfa Aesar, 99.998\%), tantalum powder (Alfa Aesar, 99.8\%), and selenium pieces (Alfa Aesar, 99.9995\%) in an evacuated silica ampoule and heating the mixture at 950$^{\circ}$ C for five days. The mixture was then cooled to room temperature, pulverized and annealed further for another 5 days at 950$^{\circ}$ C to let the precursor homogenize properly. Approximately 1.5 g of the resulting powder was combined with 0.28 g of iodine and placed in a fused silica tube of 14 mm inner diameter. The tube was evacuated and sealed under vacuum. The sealed ampoule, 10 cm in length, was placed horizontally in a single-zone furnace with the hot zone (furnace center) maintained at 940$^{\circ}$ C for two weeks. This process yielded multiple well-faceted, flat, plate-like single crystals of Co$_{1/4}$TaSe$_2$ in the hot end of the tube.\\

\noindent\textbf{Synchrotron-based high-resolution ARPES measurements}\\

\noindent Synchrotron-based high-resolution ARPES measurements on Co$_{1/4}$TaSe$_2$ were performed at the Stanford Synchrotron Radiation Lightsource (SSRL) Endstation 5-2. Millimeter-sized single crystals were prepared by mounting onto the metal sample holder and sandwiched under a ceramic post held together by silver epoxy paste. The sample was cleaved at a temperature of 7 K and with a pressure better than 5$\times$10$^{-11}$ Torr maintained during the course of the measurements. The energy and angular resolution was set to 20 meV and 0.1$^{\circ}$, respectively.\\

\noindent\textbf{Synchrotron-based Spin-Resolved ARPES measurements}\\

\noindent Synchrotron-based Spin-Resolved ARPES measurements on Co$_{1/4}$TaSe$_2$ were performed at the Advanced Light Source (ALS) Beamline 10.0.1.2. These measurements utilized a Scienta DA30 hemispherical analyzer equipped with two Focus FERRUM VLEED spin polarization detectors. The Sherman function value for the Ferrum VLEED detectors is 0.18.  The sample was cleaved at a temperature of 7 K and with a pressure better than 5$\times$10$^{-11}$ Torr maintained during the course of the measurements. The energy and angular resolution was set to 40 meV and 0.5$^{\circ}$, respectively.\\

\noindent\textbf{Ab initio density functional theory calculations}\\

\noindent First-principles Density Functional Theory (DFT) calculations were performed using a plane-wave basis set within the pseudo-potential framework, as prescribed in the Vienna Ab initio Simulation Package (VASP) \cite{Kresse_moldyn,Kresse_VASP}. The exchange-correlation interactions were treated using the Generalized Gradient Approximation (GGA) within the Perdew-Burke-Ernzerhof (PBE) functional \cite{PBE_EX}. A plane-wave energy cutoff of 600 eV was employed throughout the calculations. Structural optimization was conducted by relaxing the internal atomic positions until the Hellmann-Feynman forces were reduced below 0.001 eV/$\text{\AA}$, while keeping the lattice parameters fixed at experimentally determined values (a=b=6.88 $\text{\AA}$ and c = 12.47 $\text{\AA}$). A 9$\times$9$\times$4 $k$-point mesh was utilized for Brillouin zone sampling, and the electronic self-consistency criterion was set to 10$^{-9}$ eV.\\

\noindent\textbf{\large{Data Availability}}\\
\noindent The first-principles relaxed CIF atomic coordinates and ARPES source data for the Fermi surface, the dispersion along the $\text{M}'-\Gamma'-\text{M}'$ for both 7 K and 200 K, and the spin-resolved EDCs along the $\Gamma'-\text{M}'$ direction are available in the UCF STARS database, at \href{https://stars.library.ucf.edu/datasets/30}{\textcolor{blue}{https://stars.library.ucf.edu/datasets/30}}. The data supporting the findings of this study are available within the paper and the Supplementary Information. Further data are available from the corresponding author upon request.\\

\pagebreak

\raggedbottom

\noindent\textbf{\large{Acknowledgments}}

We acknowledge Makoto Hashimoto and Donghui Lu for their assistance at beamline 5-2 of the Stanford Synchrotron Radiation Lightsource (SSRL). Use of SSRL, located at SLAC National Accelerator Laboratory, is supported by the U.S. Department of Energy (DOE), Office of Science, Office of Basic Energy Sciences, under Contract No. DE-AC02-76SF00515. We also thank Sung-Kwan Mo and Alexei Fedorov for beamline assistance at the Advanced Light Source (ALS), Lawrence Berkeley National Laboratory. The Advanced Light Source at Lawrence Berkeley National Laboratory (LBNL), which is supported under the DOE Contract No. DE-AC02-05CH11231. The work at Hiroshima University was supported by JST EXPERTJ, Japan Grant Number JPMJEX2510. \\

\noindent\textbf{\large{Funding Statements}}

\noindent Work performed by M.N. at UCF was supported by the DOE Office of Science, Basic Energy Sciences (BES), under Award DE-SC0024304. N.J.G., R.B.R., and I.I.M. were supported by the Army Research Office under Cooperative Agreement Number W911NF-22-2-0173.\\

\noindent\textbf{\large{Author Contributions}}

\noindent M.N., N.J.G., and I.I.M conceived the project. R.B.R. and N.J.G. synthesized the crystals and studied the bulk resistivity and magnetic properties. M.S., M.I.M., A.P.S., A.K.K., H.S., and M.N. conducted ARPES measurements. S.S. and I.I.M. performed DFT calculations. M.S. and M.N. wrote the paper with contributions from all authors. M.N. was responsible for the overall research direction, planning, and integration among different research units.\\

\setcounter{figure}{0}
\renewcommand{\thefigure}{{S\textbf{\arabic{figure}}}}

\pagebreak

\title{Supplemental Information: Observation of Altermagnetic Spin-Splitting in an Intercalated Transition Metal Dichalcogenide}%

\maketitle

\clearpage

\section{Supplementary Note 1. Magnetoresistance and Heat Capacity Measurements}

Supplementary Figure 1a shows the magnetoresistance of Co$_{1/4}$TaSe$_2$ measured with current in an arbitrary
direction in the ab-plane and magnetic field along the c-axis. The magnetoresistance is defined
as MR = $(\rho(H)-\rho(H=0))/\rho(H=0)$. At 1.8 K the MR at 9 T is very small (0.9\%), which is consistent with those observed in other members of the intercalated TMDs \cite{CoNbSe_Regmi_Intro, Ghimire_CoNb3S6}. The MR decreases with increasing temperature, attaining a value of 0.014\% at 200 K. The temperature-dependence of the heat capacity measured on Co$_{1/4}$TaSe$_2$ single crystals is presented in Supplementary Fig. 1b. For temperatures between 0 and 178 the heat capacity shows a Debye-like monotonic increase with increasing temperature. At T = 178 K a clear lambda-like anomaly is present, indicating the presence of a second-order phase transition, consistent with that reported in a previously \cite{MandujanoCoTaSeMagnetic}.

\section{Supplementary Note 2. Core-level Photoemission Spectrum}

Supplementary Figure 2 presents the core-level spectrum for the Co$_{1/4}$TaSe$_2$ sample at the measurement position for the ARPES presented in the main text. A photon energy of 120 eV was used for the collection of the core-level spectrum. Prominent SOC-split Ta 4$f$ and Se 3$d$ peaks were observed at $E-E_F=21.6~eV$, 23.5 eV, 52.7 eV, and 53.5 eV, respectively. Surface replicates were observed for Se peaks at 53.1 and 52.3 eV, indicating that a Se-plane cleavage was measured. These results are in agreement with micro-ARPES measurements on closely related Co$_{1/4}$NbSe$_2$, which demonstrated the prevalence of the Se-cleavage \cite{Vita_Core-Level}.

\section{Supplementary Note 3. Momentum Dependence of Temperature-Dependent Electronic Band Reconfiguration}

Supplementary Figure 3 presents the ARPES-measured dispersion cuts along the $\overline{\Gamma}-\overline{\text{K}}$ direction (within the $\Gamma$-K-H-A plane). The apparent disappearance of the petal-pockets from the Fermi surface in Fig. 3b of the main text can be explored by considering the dispersion cuts along the $\overline{\Gamma}-\overline{\text{K}}$ direction. The front of the petal-pocket is formed by the band crossing the Fermi energy closest to the $\overline{\Gamma}$ point, which can be clearly seen as a sharp feature in the top row of Supplementary Fig 3. However, the high-temperature dispersion results, taken at a temperature of 200 K, does not show a removal of the band but a suppression of its intensity near the Fermi energy.

\section{Supplementary Note 4. Photon Energy Variation of the Fermi Surface}

Supplementary Figures 4-6 present the photon-energy variation of the obtained ARPES spectrum for Co$_{1/4}$TaSe$_2$. Supplementary Figure 4a-e presents the in-plane Fermi surfaces measured using 40, 45, 50, 55, and 60 eV photons, respectively. Supplementary Fig. 4f presents the Fermi surface within the $\Gamma$-M-L-A plane (See Fig 1c of the main text), as obtained by varying the photon energy. The main circular intensity shows little photon-energy dependence, indicating that they form quasi-2D pockets. The petal-pockets show more pronounced $k_z$ dependence, as can be seen by the pocket moving closer, then further from the $\overline{\Gamma}$ (forming the $\Gamma$-A line) and the narrowing/broadening of the petal pocket width. The presence of open Fermi surface pockets with weak $k_z$ dispersion is a signature of quasi-2D systems, where electrons are tightly confined within atomic layers (in-plane) and weakly coupled between them (out-of-plane). The layered crystal structure naturally leads to anisotropic orbital overlap, with significantly reduced dispersion along $k_z$. The modest $k_z$-dependence of the main pocket suggests finite but weak interlayer coupling, while the petal features are highly 2D and likely tied to symmetry or hybridization effects. Supplementary Figure 5 further highlights this trend by presenting the low-energy CECs, taken at $E-E_F=$ 0 meV, -25 meV, and -50 meV.

The Fermi surfaces and CECs taken within the $\Gamma$-M-L-A plane are presented in Supplementary Figure 6. The top row presents the CECs for both low temperature (T = 7 K), while the bottom row shows the high temperature (T = 200 K) results. Weakly dispersing, yet clearly three-dimensional dispersion is demonstrated for each of the pockets, which appear to gently oscillate with increasing $k_{\perp}$.

\begin{figure*}
	\centering
	\includegraphics[width=0.9\linewidth]{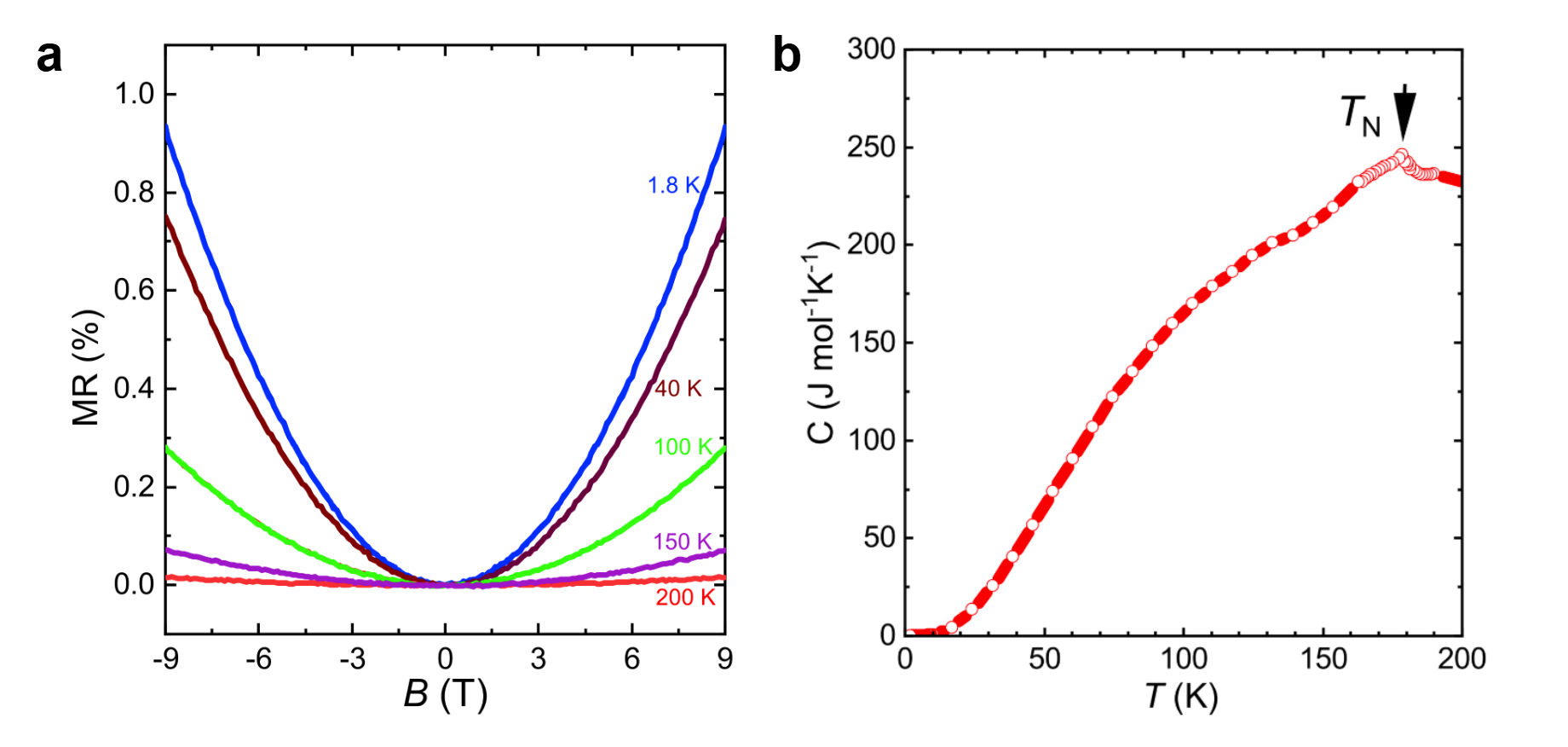}
	\caption{\textbf{Magnetoresistance and Heat Capacity of Co$_{1/4}$TaSe$_2$ single crystals.} \textbf{a} The magnetoresistance is measured with the current applied within the ab plane and a magnetic field applied perpendicularly along the c-axis. Sample temperatures of 1.8 K (blue curve), 40 K (brown), 100 K (green), 150 K (purple), and 200 K (red) are indicated by their respective colored curves. \textbf{b} The heat capacity vs temperature measured under zero applied magnetic field. The downward-pointing arrow indicates the N\'eel temperature (T$_{\text{N}}$=178 K).}
	\label{fig:mr-rr1-179a-cota4se8-h-par-c-i-par-ab}
\end{figure*}

\begin{figure*}
	\centering
	\includegraphics[width=1\linewidth]{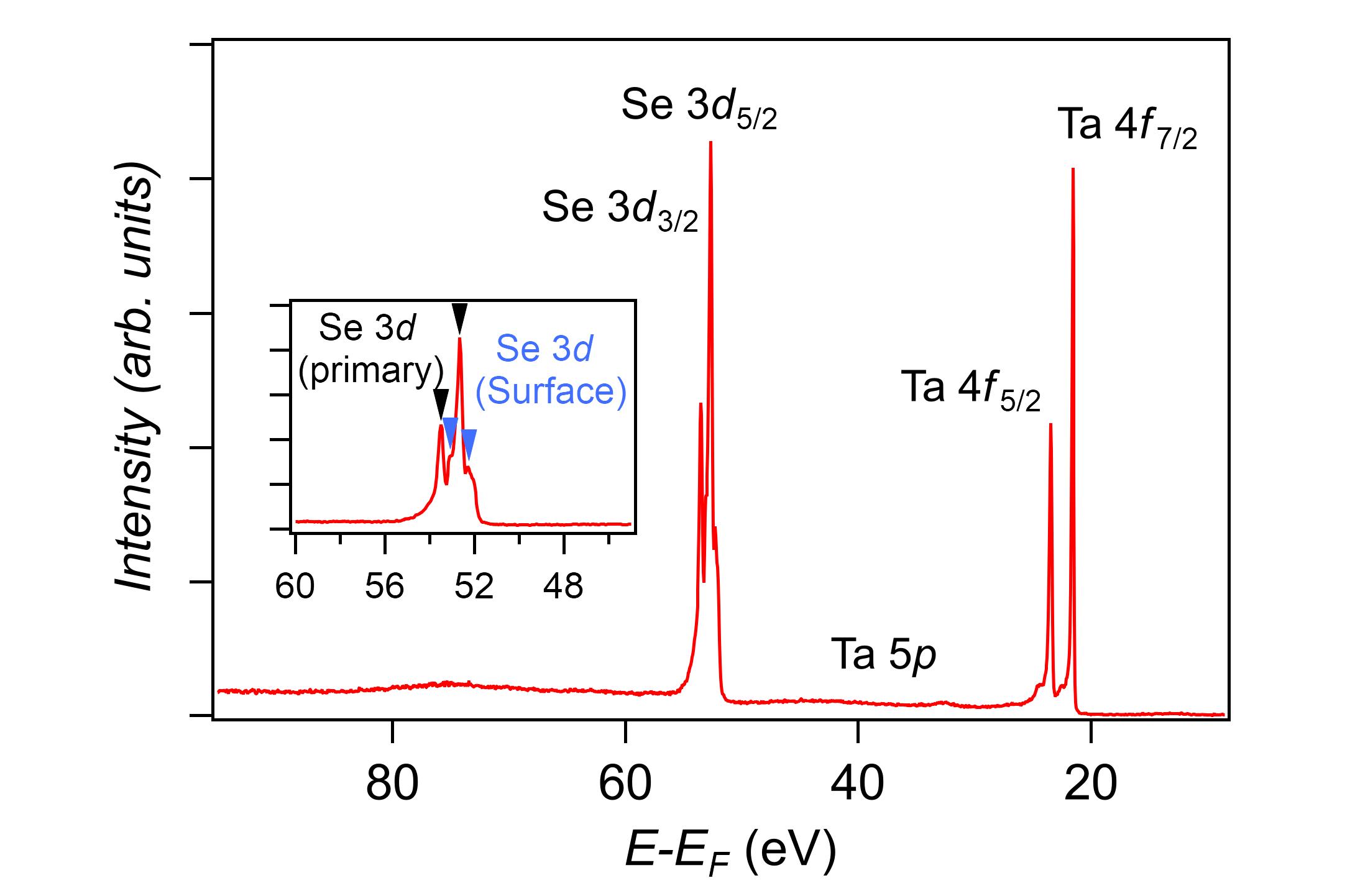}
	\caption{\textbf{Core-level spectrum for Co$_{1/4}$TaSe$_2$}. Peaks associated with Ta 4\textit{f}, Ta 5\textit{p}, and Se 3\textit{d} orbitals are identified, and labels are placed above the corresponding features. Spin-orbit split orbitals are labeled by their orbital angular momenta. The inset presents a zoomed view of the Se core levels. The primary bulk spin-orbit-split peaks are indicated by black arrows, while duplicate peaks associated with the surface environment are indicated by blue arrows.}
	\label{fig:layout8}
\end{figure*}

\begin{figure*}
	\centering
	\includegraphics[width=1\linewidth]{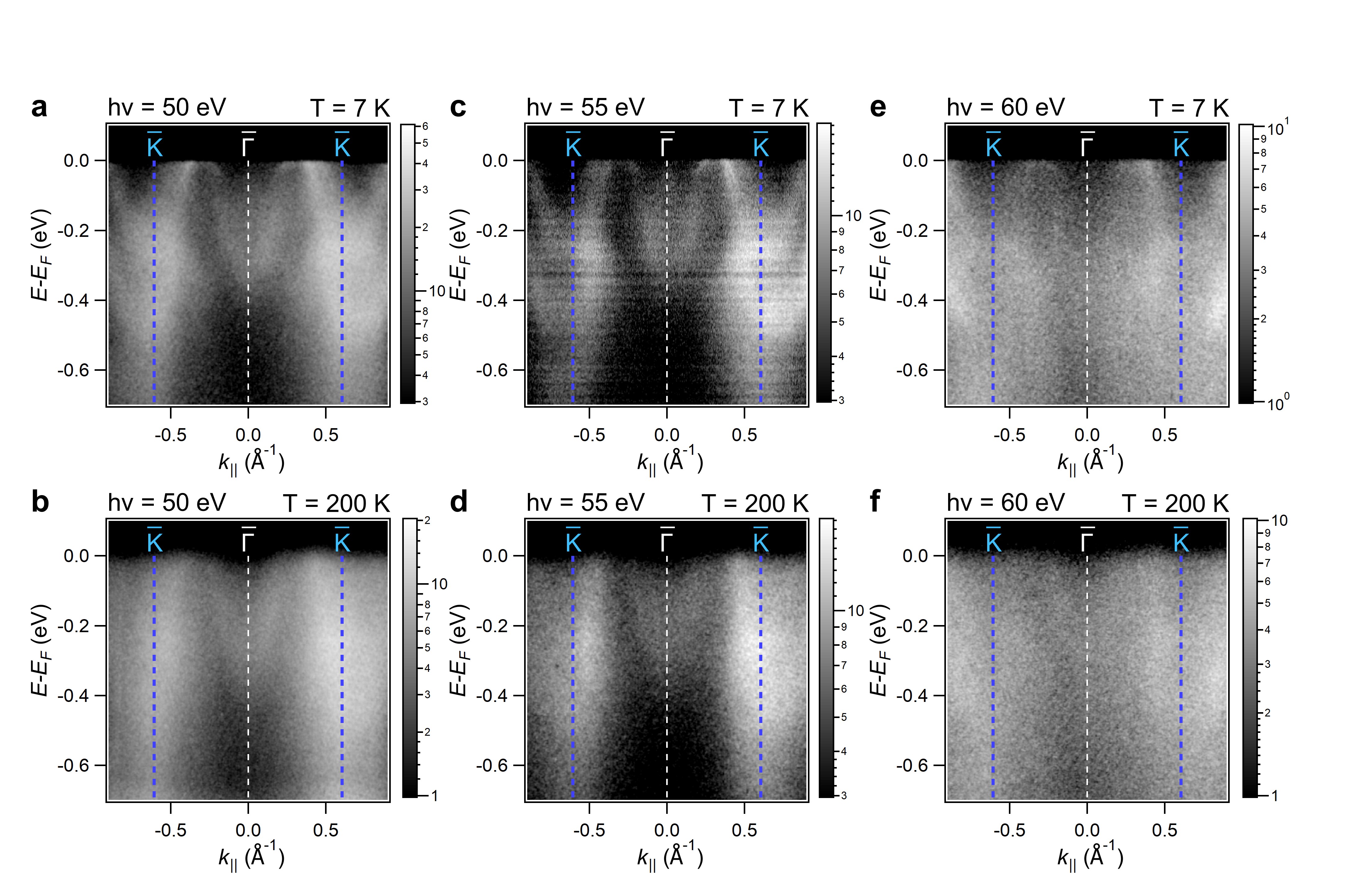}
	\caption{\textbf{
    Temperature-dependence along $\overline{\Gamma}$-$\overline{\text{K}}$.} \textbf{a,b} present the ARPES-measured dispersion cuts along the $\overline{\Gamma}$-$\overline{\text{K}}$ using 50 eV photons. \textbf{a} shows the low-temperature dispersion cut and \textbf{b} shows the high-temperature measurement result. \textbf{c,d} Presents the same information using 55 eV photons, and \textbf{e,f} does so using 60 eV photons. Here overline bars are used over the high-symmetry labels to indicate they are located at an arbitrary $k_z$.}
	\label{fig:layout8}
\end{figure*}

\begin{figure*}
	\centering
	\includegraphics[width=1\linewidth]{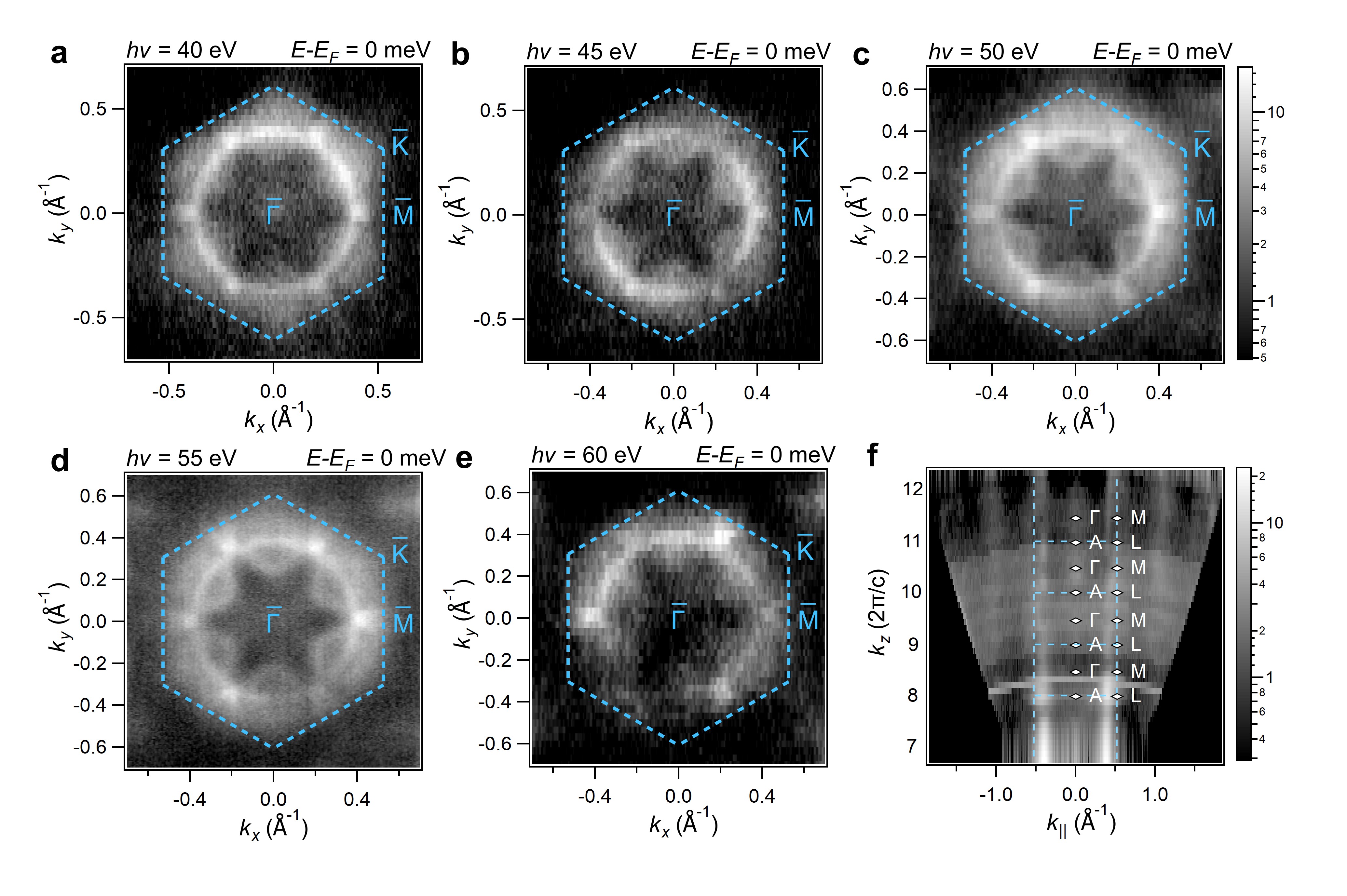}
	\caption{\textbf{Variation of ARPES-measured Fermi surface with photon energy.} Experimental Fermi surfaces were measured using (a) $h\nu$ = 40 eV, (b) 45 eV, (c) 50 eV, (d) 55 eV, and (e) 60 eV photons. (f) Fermi surface parallel to the $\overline{\Gamma}-\overline{\text{M}}$ direction along the out-of-plane momentum. The vertical axis and dashed red horizontal lines correspond to the 7-12th BZ $\Gamma$ point. The two vertical dashed lines label the $\overline{\text{M}}$ point projection. The out-of-plane Fermi surface was obtained using 30-140 eV photons. Here overline bars are used over the high-symmetry labels to indicate they are located at an arbitrary $k_z$. Panels (a-c) are plotted with intensity scaling indicated by the intensity bar on the right of panel (c), while panel (f) has intensity indicated by the scale on the right.}
	\label{fig:layout3}
\end{figure*}

\begin{figure*}
	\centering
	\includegraphics[width=1\linewidth]{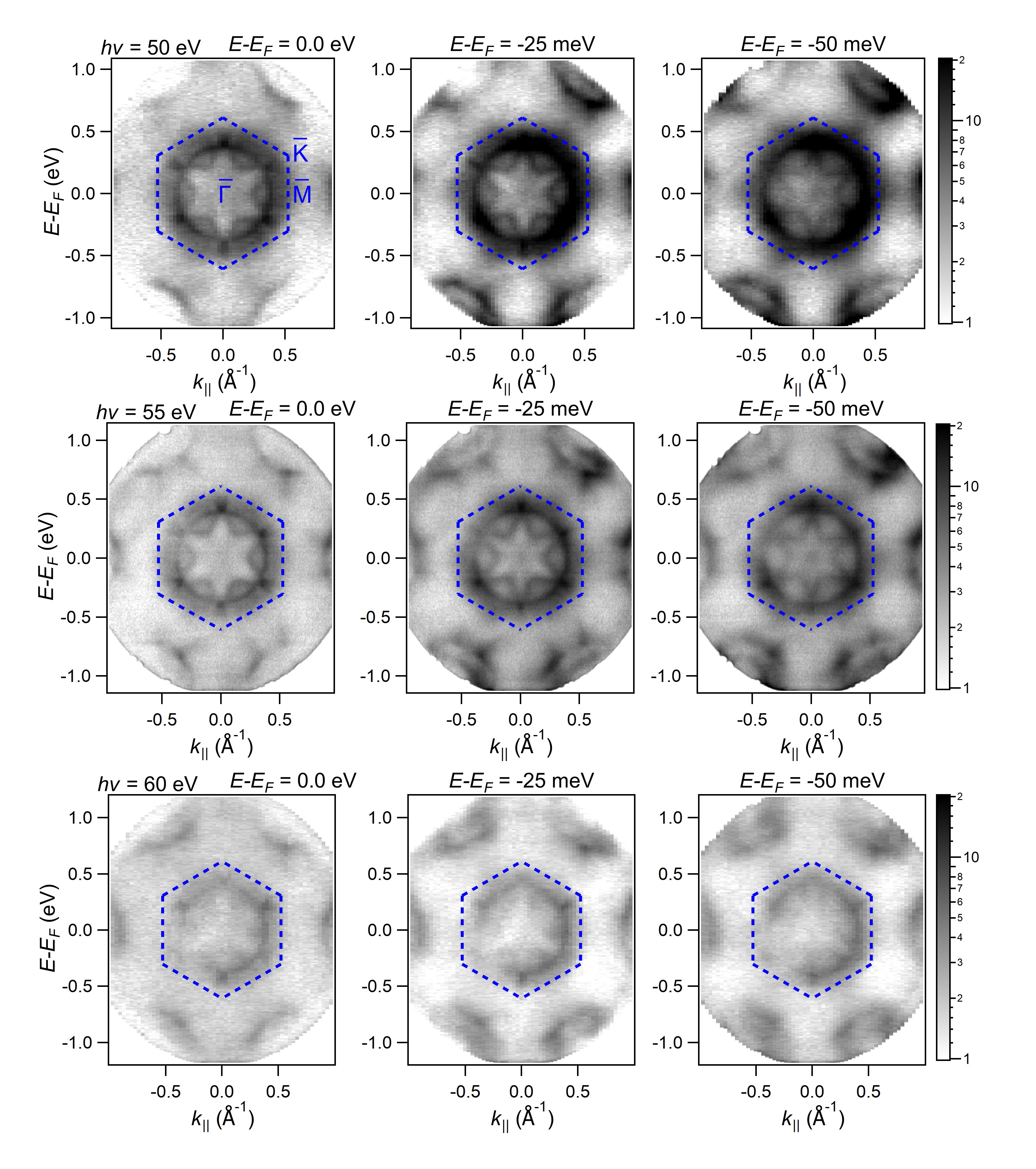}
	\caption{\textbf{Fermi surface and low-energy CECs using different photon energies.} The rows present the CECs measured using photon energies of 50 eV, 55 eV, and 60 eV, going from top to bottom. The columns present the CECs taken at the Fermi surface, $E-E_F$ = -25 meV, and -50 meV, going left to right. Each panel contains an outline of the first surface-projected BZ, with high-symmetry labels included on the top-left panel. Intensity bars are included on the far right panels. All panels share the same intensity scaling.}
	\label{fig:layout3}
\end{figure*}

\begin{figure*}
	\centering
	\includegraphics[width=1\linewidth]{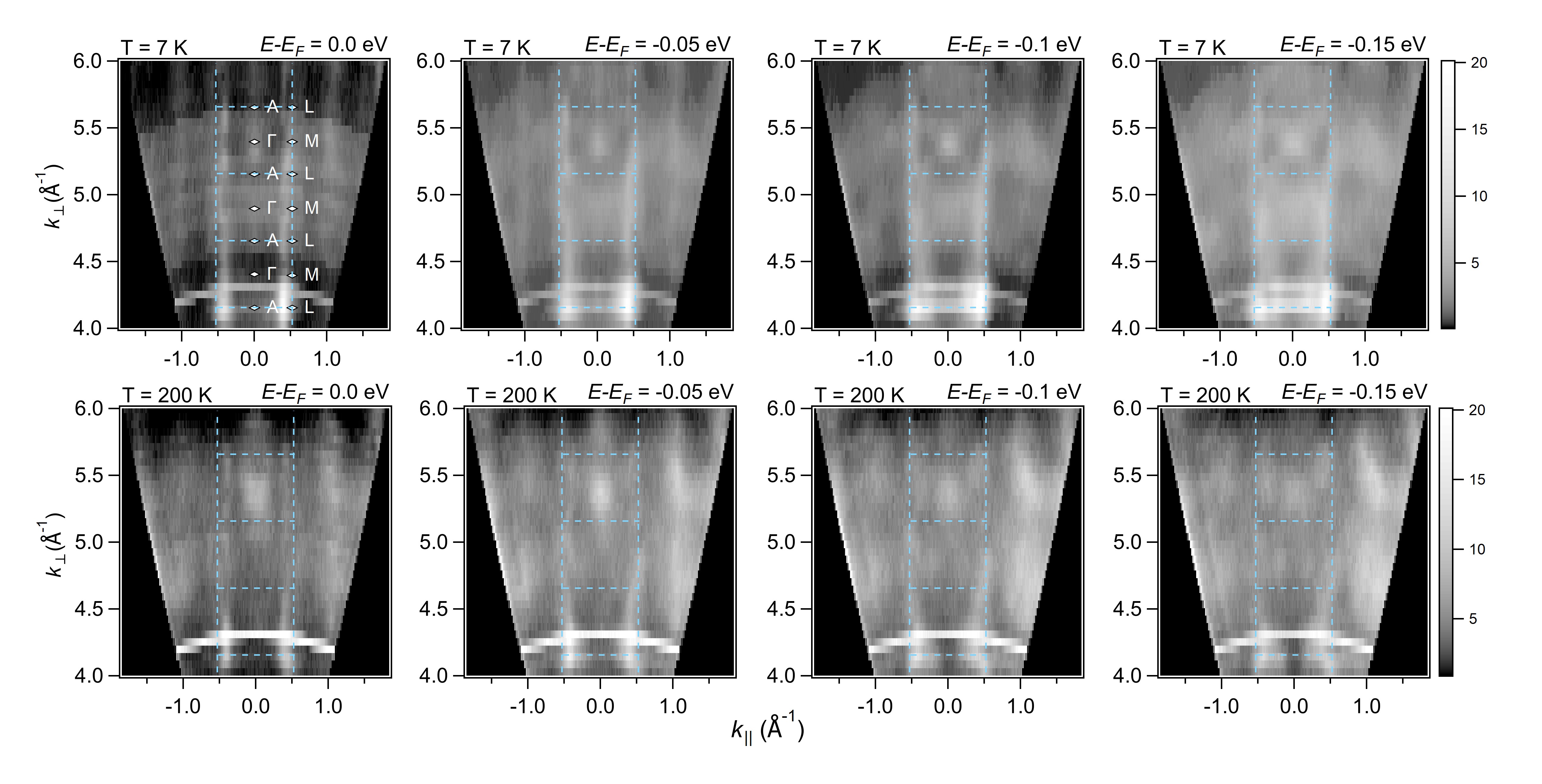}
	\caption{
    \textbf{Temperature dependence of the Fermi surface and CECs along the out-of-plane momentum direction.} The top row shows the low-temperature (T = 7 K) CECs at $E-E_F$ = 0.0 eV, -0.05 eV, -0.1 eV, and -0.15 eV, going left to right. The bottom row shows the same CECs, but measured at a higher sample temperature of T = 200 K, above the N\'eel temperature. In each panel, the rectangular projection of the bulk BZ is indicated by the blue dashed line. The top-left panel includes the high-symmetry labels. Intensity bars are included on the far right panels. All panels share the same intensity scaling.
    }
	\label{fig:layout3}
\end{figure*}

\end{document}